# $p$-FP: Extraction, Classification, and Prediction of Website Fingerprints with Deep Learning

Se Eun Oh
*University of Minnesota*

Saikrishna Sunkam
*University of Minnesota*

Nicholas Hopper
*University of Minnesota*

## Abstract

Recent advances in learning Deep Neural Network (DNN) architectures have received a great deal of attention due to their ability to outperform state-of-the-art classifiers across a wide range of applications, with little or no feature engineering. In this paper, we broadly study the applicability of deep learning to website fingerprinting. We show that unsupervised DNNs can be used to extract low-dimensional feature vectors that improve the performance of state-of-the-art website fingerprinting attacks. When used as classifiers, we show that they can match or exceed performance of existing attacks across a range of application scenarios, including fingerprinting Tor website traces, fingerprinting search engine queries over Tor, defeating fingerprinting defenses, and fingerprinting TLS-encrypted websites. Finally, we show that DNNs can be used to predict the fingerprintability of a website based on its contents, achieving 99% accuracy on a data set of 4500 website downloads.

## 1 Introduction

Website Fingerprinting (WF) refers to a network-level traffic analysis attack in which the timing, direction, and volume characteristics of encrypted traffic between a web client and a proxy are used to identify the websites visited by the user, or to attempt to differentiate between visits to *monitored* (sites that might be targeted for censorship) or *unmonitored* sites. First introduced by Hintz in 2002 [16], recent work has focused primarily on the application of WF attacks to perhaps the most widely-used anonymity network, Tor. In this context, the attacker monitors the traffic between a Tor client and the guard relay, extracts features from this traffic, and then attempts to classify this traffic based on these features.

Security researchers have adopted a variety of machine learning algorithms for WF attacks on Tor. Wang *et al.* [38] utilized k-Nearest Neighbors ($k$-NN) with a new weight learning technique, Panchenko *et al.* [28] proposed a scalable attack with Support Vector Machine (SVM), and Hayes and Danezis [13] adapted Random Forest classification ($k$-FP) to be resilient to WF defenses and noise. These classifiers have attained high true positive rates (TPR) and low false positive rates (FPR) at the cost of a great deal of human effort in feature engineering and clever algorithm adaptations. The primary contribution of each of these works was to improve the performance of classifiers by introducing new classifiers or discovering new feature sets.

Recently, Deep Neural Networks (DNN) have achieved impressive results in diverse research areas such as image recognition [14, 22, 34], game playing [37], and speech recognition [35]. Because DNNs can fit arbitrary functions with little prior knowledge and can also use multiple layers to learn different levels of abstraction of input data [6], they have the potential to automate both types of contribution to the literature on WF attacks.

There have been several efforts to apply DNN to traffic analysis. Wang [41] adopted DNN for protocol detection such as SSL. Schuster *et al.* [33] applied DNN to encrypted video stream analysis to identify which streaming videos a user watched. Rimmer *et al.* [31] revisited WF with deep learning algorithms using an immense dataset. However, these works did not attempt to isolate the benefits of automated feature engineering from classifier improvements, and primarily consider the closed world or binary classification settings.

In this paper, we extensively explore DNN to propose a variety of applications to traffic analysis, which were not covered in previous work [31].

**Contributions.** We summarize key findings as follows:

**Various Fingerprinting Attacks.** We studied the suitability of Multilayer Perceptrons (MLP) and Convolutional Neural Networks (CNN)[1] across a wider range

---

[1]The MLP architecture features several layers of neurons, each fully connected to the layer before it; a CNN convolves its input with sev-

of settings than Rimmer *et al.*: in addition to identifying top Alexa web sites in the closed and binary open-world setting, we also study the performance of these architectures in other WF tasks including multi-class open-world classification, search query (keyword) fingerprinting [26], Onion Service fingerprinting, TLS-encrypted website fingerprinting, and WF against four traffic padding schemes – BuFLO [11], Tamaraw [10], WTF-PAD [18] and Walkie-Talkie [40]. We show that for all of these tasks, DNNs can achieve equivalent or better results to those published in the literature.

Overall, CNN is a better model across various scenarios and in particular, it is more effective against WF defenses. CNN is capable of identifying 100 monitored websites against 100,000 unmonitored websites with 95% true positive rate and 0.04% false positive rate, using only 9,000 traces of monitored training data. In particular, BuFLO, WTF-PAD, and Walkie-Talkie with previously recommended parameters become less effective against CNN since we successfully conduct the multiclass classification on 100 websites with 45%, 58%, and 54% accuracy, respectively.

**Feature Engineering.** To isolate the benefits of DNNs as feature extractors versus classifiers, we also investigate the use of an autoencoder (AE), an unsupervised learning technique to extract low-dimensional representations of a dataset, in combination with the classifiers used in state-of-the-art WF attacks. We find that when state-of-the-art WF attacks are applied to AE-extracted features, they become more powerful than when using hand-tailored feature sets in recent WF attacks. In particular, classifier accuracy is improved using AE-extracted feature vectors with dimension as small as 20. This suggests that AEs can be a powerful technique for feature engineering in new traffic analysis attacks.

**Predicting Fingerprintability** Overdorf *et al.* conducted website fingerprintability (FP) analysis using an ensemble classifier based on three state-of-the-art WF attacks using network-level and site-level features. In this paper, we propose a novel fingerprintability predictor based on MLP using feature sets that only focus on elements in HTML documents such as statistics on links and embedded web content.

While Overdorf *et al.* adopted the F1 score of an ensemble classifier as a fingerprintability score, we use *accuracy* of a MLP classifier trained on Tor traces of a website (the proportion of correct predictions in the total number of testing instances of a website). We label HTML-specific feature vectors with 1 or 0 to indicate whether the accuracy of a website is greater than a "fingerprintability threshold" (10-90%) or not.

We predicted the fingerprintability of 4,500 instances of Alexa top 50 websites with 98-99% accuracy and less than 0.02% mean squared error (MSE) under 10-40% and 90% fingerprintability thresholds. Moreover, we find that a MLP trained on Alexa top-67 websites can successfully predict the fingerprintability of 85 low-ranked Alexa websites with 98% accuracy and 1% MSE. In addition, even when we label HTML feature vectors based on classification results by *k*-FP [13], FP classifiers show comparable performance. Our FP predictor is effective across various scales of background set, different testing datasets, and different ML algorithms. These results suggest that website designers interested in helping users avoid WF attacks (or onion service designers interested in protecting the location of their servers) can apply our classifiers to the HTML source of their sites to predict the vulnerability of content pages and alter them accordingly.

## 2 State-of-the-art Attacks

The importance of feature engineering for WF with cutting edge machine learning algorithms was first highlighted by Panchenko *et al.* [29]. Their hand-built features based on traffic volume and timing were powerful enough to have high TPR in an open world scenario.

Wang and Goldberg [39] proposed using Tor cell sequences as a new feature set and used SVM with distance based metrics to yield classifiers with 95% TPR and 0.2% FPR. The *k*-NN classifiers with revised weight learning, proposed by Wang *et al.* [38], enabled the attacker to achieve higher TPR than prior WF attacks.

Panchenko *et al.* [28] demonstrated more scalable WF with larger background datasets than previous research. They proposed CUMUL classifiers using Support Vector Machines (SVMs) with a new feature set based on cumulative traces and obtained 96-97% TPR, higher than *k*-NN classifiers [38]. Hayes and Danezis [13] conducted a thorough analysis of new categorical features and built *k*-FP classification models based on Random Forests with Hamming distance. Their classifiers were less susceptible to padding-based WF defenses such as BuFLO [11] and Tamaraw [10].

Oh *et al.* [26] extended WF with new feature sets to investigate Keyword Fingerprinting (KF) attack, which identify search engine queries over Tor that the user typed. They built svmRESP classsifiers using SVM with their RESP feature sets that focus on the response traffic to capture embedded web objects in search engine results. They achieved 83% TPR and 8% FPR with svmRESP to correctly identify 100 monitored queries against 10k unmonitored queries.

---

eral local shared "filter" units, which are locally "pooled" using a common function, before applying a single fully-connected layer; see Appendix B for further details

These works focused on finding features and machine learning models, better suited to their datasets. Their methodologies required substantial human effort to analyze features and time-consuming trial-and-error procedures to determine better classification algorithms.

Recently, Rimmer *et al.* [31] explored the application of several DNN architectures, including Stacked Denoising Autoencoder (SDAE), CNN, and Long Short-Term Memory (LSTM), to an enormous dataset to prove DNN's effectiveness and scalability for WF attacks.

In this paper, we delve into MLP, CNN, AE to explore a variety of applications including fingerprinting attacks, automated feature extraction, and website fingerprintability prediction, for traffic analysis. In particular, to show great power of automated feature engineering using unsupervised learning in DNN, we revisited *k*-NN, CUMUL, *k*-FP, and SDAE classifiers for comparison. We give more detailed background on the MLP, CNN, and AE architectures in Appendix B and describe implementation details relevant to our work in Section 4.4.

## 3 Feature Extraction

### 3.1 Dataset Overview

*Website Tor traces (WTT).* For monitored traces, we used Wang's dataset(**Wang**) [38], Tor Hidden Service dataset(**TorHS**) [13] and we collected **WTT-time** dataset consisting of 90 instances of each of Alexa top 100 websites [1]. Wang's dataset consists of 90 instances of each of 100 monitored websites, and TorHS is comprised of 90 instances of each of 30 onion services. For WTT-time dataset, we additionally downloaded HTML document files for those traffic instances to be used in Section 5. WTT-time dataset reflects the time gap, a week, between 4 different batches where each batch collects 110 traffic instances of each of 100 websites. Therefore, one batch started and the next batch executed after a week. After filtering out instances having abnormal capture files or HTML documents and sampling uniformly at random from four batches comprising remaining instances, we ended up with 90 instances for each of 100 websites. We crawlered them from June to December in 2017.

For Section 6.6, we further harvested 90 instances of 85 Alexa websites, ranked around one million. We crawled those websites in September and December, and we adopted the same batch settings as we used in WTT-time dataset collection. For background traces, we used the dataset, provided by Hayes and Danezis [13], which contains one instance of each of one million websites.

*Keyword Tor traces (KTT).* For both monitored and background traces, we used Google search query traffic instances, provided by Oh *et al.* [26], which include 100 instances of each of 100 top-ranked monitored keywords and 80,000 unmonitored instances.

*Website SSL (non Tor) traces (WST).* We harvested TLS-encrypted website traces in normal web browser settings and the dataset consists of 90 instances of each of Alexa top 100 websites for a monitored set and 9,000 Alexa websites excluding monitored websites.

### 3.2 Features for DNN

Based on various traffic dataset, we introduce features we used for MLP and CNN classifiers.

*Tor traces for WF.* Since Tor supports a specific transmission unit, known as a Tor cell (512 bytes), we extracted a Tor cell sequence, which consists of 1 and -1 indicating that the client sends 512 bytes or receives 512 bytes, respectively. In addition, since all instances should have the same feature dimension, we determined the optimal number of features using hyperparameter tuning, shown in Section 4.3.

*Tor traces for KF.* We evaluated two feature sets, Tor cell sequence and RESP trace, provided by Oh *et al.* [26]. RESP is the largest incoming burst and they extracted the sequence of cumulated payload based on TLS record length in RESP, however, we ignored the cumulative setting because it gave us lower accuracy. By applying hyperparameter tuning, we experimentally determined 2500 features as the optimal dimension and used it for all KF experiments in Section 6.

**TLS-encrypted traces.** We evaluated various feature sets including the inter-packet timing, the size of tcp packet, and the length of tls record. With packet and record size sequences, MLP did not show effective result. However, when we used the sequence of packet direction (-1 for incoming and 1 for outgoing), we reached better TPR up to 82%. In contrast, we were able to achieve 92% TPR for CNN even with features based on the sequence of the size of TLS record. Experimental details will be discussed in Section 6.3.

### 3.3 Features with Autoencoder

While an AE is widely used for reconstructing the data, it is also known for the ability of feature dimensionality reduction because an encoder projects the original feature vector into lower dimensional representation.

In this section, we explore features, learned by an AE, to show their effect on the performance of SVM, *k*-NN, and *k*-FP classifiers. In particular, this feature compression can make classifiers more efficient because their training time highly depends on the dimension of the input data, which often leads scalability issues. As shown in Figure 8c of Appendix B, we trained an encoder and a decoder and we captured feature vectors, compressed by the second hidden blue layer of an encoder. These are *encoded features*. We also varied the number of units, 10-100, in this hidden layer to compress the original features into various low-dimensional representations. We evaluated SVM, $k$-NN, and $k$-FP classifiers with encoded features and reproduced state-of-the-art WF attacks to compare their performance in Section 6.5. This shows powerful capability of an AE on learning interesting structure about the data while reducing its dimension.

With the autoencoder, we additionally tested a variational autoencoder (VAE) and extracted encoded features as explained above, however, we failed to extract meaningful traces. For VAE, we achieved 2% TPR for multiclass classification and 3% TPR for the binary decision. The reason for lower TPR with VAE was that we failed to reduce the loss, specifically KL divergence loss between latent space, which is the distribution of an encoder and its probability density function. Since this gap was huge, KL divergence blew up and further led large cost. Since the VAE injects random Gaussian noise and optimizes the likelihood, the VAE performs more nicely for datasets where the latent is important [24]. Website traffic instances are less likely to have a reasonable estimated density that the log likelihood maximization returns than the image dataset.

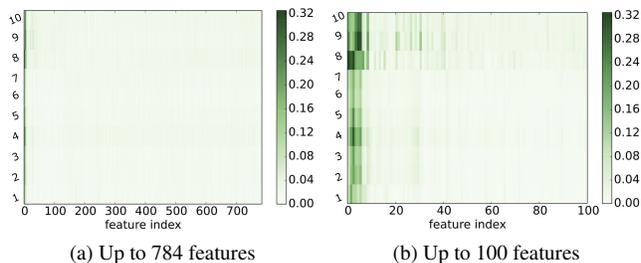

(a) Up to 784 features      (b) Up to 100 features

Figure 1: Feature importance score using LRP.

## 3.4 Feature Importance using LRP

Layer-wise relevance propagation (LRP) [4] sheds light on understanding how DNN recognizes the input dataset and further, what features in the input contribute most to DNN's output. To use LRP, first, we decompose the output of DNN into the sum of relevance scores in the input vector. To do this, we start with the relevance of the output node and iteratively backpropagate through the network. This is similar to backpropagation and finally it maps the relevance onto the input vector.

There are several decomposition methods to derive the relevance score and we chose $w^2$-rule [25]. This rule selects the closest root in direction of maximal descent when using Deep Taylor decomposition, which approximates the value of the neural network forward pass function. We map layer $l$'s activations $x_i$, where i is node at layer $l$, to node $n$'s (at layer $l+1$) activation $y_j$ and decompose the forward function in terms of $w^2$-rule to measure the relationship between $x_i$ and $y$. That is, the relevance score sent from a layer $l+1$ to its predecessor $l$, $R_{i \leftarrow j}^{(l,l+1)} = \frac{w_{ij}^2}{\sum_{i'} w_{i'j}^2} R_j^{(l)}$, where $w_{ij}$ is the weight between node $i$ at layer $l$ and $j$ at layer $l+1$.

We used LRP toolbox [20] to choose more important features in the input of Tor traffic instances and Tensorflow to build the network. We computed the relevance $R$ for each of 784 features in WTT-time dataset. After we ran 10 experiments, we did summation of $R$ for each feature index and visualize the sum score in Figure 1. The first $n$ Tor cells contribute more to the prediction output, which empowers the decision on cutting off Tor cells at the end to have the same length in data preprocessing. When we derived three $R$ scatter distributions for each of 0 (padding), 1(outgoing), and -1(incoming), which are three values in the input vector, there is no noticeable distinction between those values, which makes it hard to determine which Tor cell direction impacts most to the prediction output in terms of $R$ (Figure 8 in Appendix C).

## 4 Traffic Analysis with DNN

In this section, we introduce our DNN architectures, metrics to evaluate the performance of our classifiers, and hyperparameter tuning to find the optimal parameters to train DNN models.

### 4.1 Threat Model

We assume a network-level, passive adversary who is only able to monitor and capture network traces, sent and received by users. Since the user uses a Tor browser, the adversary is only able to observe network traffic between the client and a Tor entry guard. In addition, there are no cooperative adversaries involved in this communication (e.g.,web servers, Tor relays, etc.).

Our attacker is interested in two types of classification problems, one is to determine whether a monitored trace is in his monitored list or not (binary classification), and the other is to correctly predict which website the user visited and which keyword the user typed (multiclass classification).

### 4.2 Metrics

*Traditional metrics.* To evaluate the performance of binary and multiclass classifiers in open-world setting, we used the following metrics, adopted in previous works [13, 26, 28].

- True positive rate (TPR): This metric indicates the proportion of positive samples that are predicted as positive.

- False positive rate (FPR): This metric shows the proportion of negative samples that are mispredicted as positive.

- Bayesian detection rate (BDR): To reflect feasibility of our experiment setting, we used BDR, computed by $P(M|V) = \frac{P(V|M)P(M)}{P(V|M)P(M)+P(V|B)P(B)}$, where V means that the user is visiting one of monitored websites, $P(M)$ is the probability that the website is included in monitored set and $P(B)$ is the probability that the website is included in unmonitored set, that is $P(\neg M)$. $P(V|M)$ is computed by TPR and $P(V|B)$ is estimated by FPR. We computed BDR in a way, presented in $k$-FP attacks [13].

- Within-monitored accuracy (WMacc): In KF, we used a Within-monitored accuracy, proposed by Oh *et al.* [26] to measure the performance of multiclass classifiers. This is computed by the number of TPs, divided by the total number of monitored samples.

***Confidence threshold.*** In DNN, the output layer returns prediction probability vectors for all labels. Even though the label with the highest probability is selected as a predicted label, if its prediction probability is not high enough, the confidence in the classifier's decision becomes weakened. If the highest probability is less than the confidence threshold, we regard this case as being classified as "others". Thus, we increased the confidence thresholds, up to 90%, to show how much this factor degrades the performance of DNN since it impacts the number of FNs and TNs.

***TopK analysis.*** Based on the prediction probability vector, we consider other meaningful labels if their probabilities are high enough to be trustworthy although they are not the highest. Those labels are contained in the top $k$ list. We varied $k$, 1-5, and if there is a match between an actual label and the one in the top $k$ list, it is a TP.

In the open world experiment, if the negative (unmonitored) label is included in the top $k$ list and the actual label is positive (monitored), we always treat it as a false negative even if the top list includes a true label. We used top $k$ analysis in KF attacks(Section 6.2) and defended traffic analysis (Section 6.4).

### 4.3 Hyperparameter Tuning

Since training DNN requires hyperparameters, searching the optimal parameters is essential to ensure better prediction result. To automate this procedure, we adopted Tree of Parzen Estimators (TPE) [7], which is a form of Bayesian optimization and used `hyperopt` library [8] to implement this optimizer. First, we described the search space, as shown in Table 1, and constructed DNN using `tflearn` library [3]. Then, DNN models are trained and pass the prediction result to the optimizer, which selects optimal parameter values for the next iteration to minimize the prediction error. We summarize the search space and chosen values for each DNN model, used in our experiments, in Table 1. For feature dimension, we used 784 for website and Tor HS traces, 2,500 for keyword traces, and 10,000 for TLS-encrypted traces in MLP evaluations. In addition, we adopted 1,200 for TLS-encrypted traces and 2,500 for all other kinds of traces to measure the performance of CNN.

### 4.4 DNN Architectures

We discuss general background on DNN models in Appendix B, and details of our architectures for DNN models in this section.

*MLP.* An MLP (Figure 8a in Appendix B) consists of one input layer allowing a single vector, two hidden fully connected layers, and one output layer with softmax function. We used L2 regularizations for the first two hidden layers and the dropout between those layers to minimize the impact of overfitting on the validation of testing dataset. After hyperparameter tuning, we decided Stochastic Gradient Descent (SGD) as the best optimizer to ensure faster backpropagation. To measure the error rate of classifiers, we used the categorical cross entropy.

*CNN.* A CNN (Figure 8b in Appendix B) comprises an input layer accepting 2 dimensional data, a convolutional 2D layer with 32 filters, followed by a 2D max pooling layer, one hidden fully connected layer and an output layer with softmax function. We applied L2 regularization to the convolutional layer and the categorical cross entropy to compute the loss in classification results.

In contrast to image dataset, 2D representations do not carry meaningful information for network traffic instances since they are sequences of -1 and 1. Hence, we adopted 1 by $n$ grid for the input data format, for example, the vector of 2,500 features is represented in 1x2500, rather than 50x50, while still keeping 2D format. Moreover, we adjusted the format of filters, accordingly. For instance, we used 1x3 instead of 3x3. This transformation ensures better performance of CNN and further, helps to avoid overfitting since it reduces the degree of variation between results evaluating on different portion of dataset.

*AE.* An AE (Figure 8c in Appendix B) consists of an encoder and a decoder network and each consists of two fully connected layers. We changed the number of units in the second hidden layer to get different feature dimensions of compressed inputs to be fed into SVM, $k$-NN, and *k*-FP classifiers. To capture encoded features, we saved a trained model and retrieved weights of the second hidden layer and generated compressed vectors on target input using those weights. We used the mean squared error to calculate the loss of the classifier decisions.

## 5 Analysis on HTML documents

According to the fingerprintability analysis explored by Overdorf *et al.* [27], the fingerprintability of website traffic instances is affected by the size of websites. We aim to study the same problem, however, using DNN and HTML-specific feature set, and further discover website design features to impact the fingerprintability. We downloaded 9,000 HTML document files of Alexa top 100 websites and 7,650 files for Alexa 85 websites, ranked around 1 million to extract HTML features.

### 5.1 HTML Features

*Links and domains.* First, we fetched all links and crafted features based on the number of links and domains in an HTML DOM. In particular, we inspected how many links are from third party websites with the expectation that downloading web objects from different web servers makes the traffic pattern more various than when all downloads occur from a single web server.

*Tag path.* We built all tag paths by iteratively adding a tag if it was nested. For example, for a document consisting of <html><body><a></a><b></b></body></html>, there are 4 tag paths, <html> (depth=1), <html><body> (depth=2), <html><body><a> (depth=3), and <html><body><b> (depth=3). Based on tag paths, we extracted statistical features about the number of tags contained in a tag path, the frequency of increase or decrease in the size of tag paths, and the depth of tag paths. These features measure the degree of the complexity or simplicity of the website design.

*Tags and other elements.* We generated features about the number of tags, attributes, and comments, and the number of characters and words in `data` and `style` attributes. These features impact the size of an HTML DOM as well as the website.

*Embedded files.* Different types of files are embedded in an HTML document file and traffic associated with fetching those resources makes website more vulnerable against WF attacks. We found that all image and video contents are nested in an `img` tag. Based on this fact, we computed the number and the proportion of image and video files and furthermore, we located specific file extensions to obtain counts and proportions of them (e.g., jpg, gif, ico, html, etc.). They impact the size of websites.

*Other features.* We computed the total transmission time based on the start and end time in a traffic capture file (e.g., pcap), and the size of a capture files and HTML document files.

All features resulted in total 65 features, named *HTML features*, and we generated the rank for each feature by looking at each column in the matrix. For example, if we have 3 instances, [[3,19,...,10], [7,10,...,201], [17,7,...,25]], we converted features into the rank information, [[1,3,...,1], [2,2,...,3], [3,1,...,2]] and fed them into classifiers to determine whether a website is fingerprintable or not. We listed 65 features in Appendix D.

### 5.2 Fingerprintability Classifier.

*Fingerprintability (FP) thresholds.* After training and testing MLP classifiers with 9,000 monitored website traffic instances and 20k-100k background instances, we computed the proportion of correct predictions for each monitored website. We named it *accuracy* of each website. Rather than restricting the accuracy threshold to a certain cutoff, we applied 9 different thresholds, 10-90%, to determine if it is fingerprintable. For example, if the threshold is 10%, classifiers decide whether the accuracy of the website is less than 10% or not. If yes, it is not fingerprintable.

*Metrics.* HTML features of those website traffic instances were labeled with 1 or 0, where it is 1 if its accuracy was greater than a FP threshold, otherwise, it is 0. This labeling led unbalanced number of instances for each class, as shown in Table 2. To overcome the imbalance in the dataset, we computed a total accuracy and a mean squared error (MSE) by applying class weights, derived based on the size of training samples for different FP thresholds (Table 2). That is, $MSE(y,y') = \frac{1}{N}\sum_{i=0}^{N} w_i(y-y')^2$ and $Accuracy(y,y') = \frac{1}{N}\sum_{i=0}^{N} w_i 1(y=y')$, where $y$ and $y'$ are a true and a predicted value, $w_i$ is the weight of a sample $i$, and $N$ is the total number of samples.

We used 4,500 instances of 50 websites for training dataset and 4,500 instances of other 50 websites for testing dataset [2].

*Gini importance.* To measure the feature importance score for each of 65 features, we used Gini importance [9], which is derived as the total number of splits including target feature over the total number of samples that the feature splits, by running Random forest classifiers. In other words, this score means an average purity, earned by splits of target feature. We used Gini importance to see which HTML DOM features contributed most to the fingerprintability of websites in Section 6.6.

With FP thresholds and metrics, we adopted MLP to evaluate its applicability to the fingerprintability pre-diction. All experimental results are discussed in Section 6.6.

---
[2]The fact that all features have the same dimension, 65 in all 9,000 instances can relieve the concern on overfitting even though we only used 4,500 instances for training.

Table 1: DNN Hyperparameter tuning using HyperOpt.

| DNN | MLP | | AE | | CNN | |
|---|---|---|---|---|---|---|
| **HyperParam** | **Choice** | **Space** | **Choice** | **Space** | **Choice** | **Space** |
| input dim | 784,2500,10k | 0∼10k | 784 | 0∼5000 | 1200,2500 | 700∼5000 |
| optimizer | SGD | SGD,Adam | Adam | SGD,Adam | SGD | SGD,Adam,RMSProp |
| learning rate | 0.08∼0.09 | 0.001∼0.1 | 0.001 | 0.001∼0.1 | 0.1 | 0.001∼0.1 |
| epoch | ≤50 | 10∼1000 | 10 | 10∼1000 | 50∼100 | 10∼1000 |
| batch | ≤40 | 10∼100 | 256 | 10∼300 | ≤30 | 28∼128 |
| number of layers | 4 | 3∼5 | 4 | 3∼5 | 5 | 3∼7 |
| hidden units | 600∼700 | 10∼1000 | 200∼300 | 10∼1000 | 200∼300 | 10∼1000 |
| dropout | 0.8 | 0.2∼0.9 | - | - | 0.8 | 0.2∼0.9 |
| activation | tanh | tanh, relu, sigmoid | tanh | tanh, relu | tanh | tanh, leaky-relu, elu |
| number of filters | - | - | - | - | 32 | 4∼128 |
| filter size | - | - | - | - | 3 | 2∼16 |
| kernel size | - | - | - | - | 2 | 2∼50 |

Table 2: Number of instances of less than and greater than a FP threshold in 9,000 web trace instances.

| FP thr | # less | # greater |
|---|---|---|
| 10 | 68 | 8932 |
| 20 | 316 | 8684 |
| 30 | 680 | 8320 |
| 40 | 972 | 8028 |
| 50 | 2161 | 6839 |
| 60 | 3399 | 5601 |
| 70 | 5222 | 3778 |
| 80 | 6787 | 2213 |
| 90 | 8036 | 964 |

## 6 Experiment

In this section, we present the results of our experiments applying DNN models to website fingerprinting in the range of scenarios described previously. In each scenario, the attacker either tries to decide whether the user was browsing a monitored website or not (binary classification) or which website the user visited (multiclass classification),

We explored using DNNs for WF attacks on Tor (Section 6.1), KF attacks on Tor (Section 6.2), WF attacks on TLS-encrypted traffic (Section 6.3), WF attacks on Tor with padding defenses Section 6.4), feature engineering for state-of-the-art attacks (Section 6.5), and fingerprintability prediction (Section 6.6), and report the best experimental setting for each scenario.
*Experiment setup.* We used Tensorflow [2] with TFLearn [3] front end for the implementation of DNN classifiers. We split the WTT, KTT, and WST datasets into training and testing datasets by the ratio, 60:40. We constructed 20 different iterations, where each iteration consists of randomly chosen monitored and background instances. We randomly selected the same number of instance indices for each monitored website to ensure that the number of classes in each iteration was equal. We adopted 20 iterations for all experiments in this section.

We used 8 cores and 32GB of memory for the experiments using MLP and AE models and other machine learning algorithms, and 32 cores and 258GB of memory for CNN classifiers. With those resources, the longest job was finished within 2 days. With GPUs, this running time would be considerably reduced.

### 6.1 Website Traffic Analysis.

We evaluated both MLP and CNN classifiers using WTT-time and Tor HS datasets in the open world setting by varying different factors.
*Background set.* To show the effect of background set, we trained MLP and CNN classifiers with the WTT dataset, where we have 9,000 traffic instances of 100 monitored websites (each instance labelled according to its site) and 1 traffic instance of 20,000, 50,000, and 100,000 background websites (all labelled with a single "background" label).

Increasing the number of unmonitored traffic instances weakened the performance of both classifiers but not significantly, since we achieved 90% TPR and 1% FPR for MLP multiclass classifiers even against 100k background dataset, given BDRs ranging from 87-89% (Table 4). We conclude that DNNs are very successful at open-world WF attacks, as shown in Tables 3 and 4.
*Confidence threshold.* To investigate the reliability of decisions made by DNN classifiers, we additionally ap-plied 0-90% confidence thresholds (0 corresponds to argmax) to prediction probabilities to decide if a classifier gives confident TPs. After we evaluated MLP and CNN classifiers using 9,000 monitored traces and 20k background traces, increasing the confidence threshold reduced the number of confident TPs, which lowered TPR, increased the number of TNs, which decreased FPR, and elevated BDR (Figure 8 and 3).

Table 3: WF with MLP by varying the size of background sets (TPR(T), FPR(F), and BDR(B) (%), and H=Tor HS)

| Size | Multiclass | | | Binary | | |
|---|---|---|---|---|---|---|
| | T | F | B | T | F | B |
| 20k | 94±1 | 5±1 | 89±1 | 95±1 | 0.3 | 99.2 |
| 50k | 92±1 | 2 | 88±1 | 93±1 | 0.1 | 99.2 |
| 100k | 90±1 | 1 | 87±1 | 91±1 | 0.07 | 99.1 |
| 30k(H) | 94±2 | 3 | 70±1 | 96±1 | 0.06 | 97.1 |

Table 4: WF with CNN by varying the size of background sets (TPR(T), FPR(F), and BDR(B) (%), and H=Tor HS)

| Size | Multiclass | | | Binary | | |
|---|---|---|---|---|---|---|
| | T | F | B | T | F | B |
| 20k | 97±1 | 6±1 | 88±1 | 98±1 | 0.3 | 99.3 |
| 50k | 95±1 | 3±1 | 85±2 | 96±1 | 0.1 | 99.1 |
| 100k | 94±2 | 2±1 | 84±2 | 94±2 | 0.04 | 99.5 |
| 20k(H) | 98.49 | 3.69 | 78±1 | 98.91 | 0.18 | 98.6 |

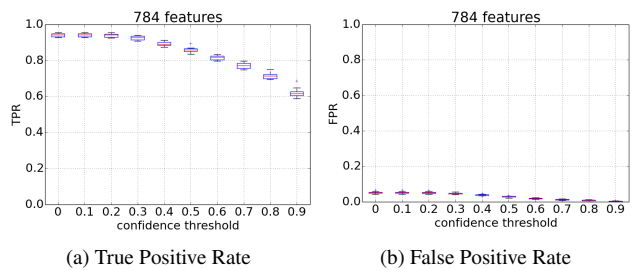

(a) True Positive Rate  (b) False Positive Rate

Figure 2: WF evaluation using MLP in the multiclass setting by varying confidence threshold and unmonitored set.

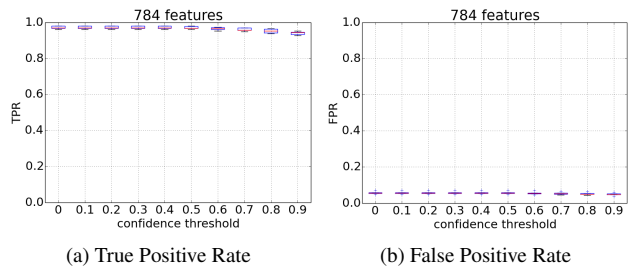

(a) True Positive Rate  (b) False Positive Rate

Figure 3: WF evaluation using CNN in the multiclass setting by varying confidence threshold and unmonitored set.

*Network.* As shown in Table 3 and 4, CNN classifiers showed better results except FPR in multiclass classification, although the gap between FPRs of both classifiers was not high. More interestingly, as shown in Figure 3, the increase in the confidence threshold hardly impacted the performance of classifiers across all metrics.

*Summary.* Feeding more background instances somewhat decreases the performance of DNN classifiers, however, CNN models demonstrate very high multinomial classification accuracy with 92% TPR and 1% FPR against 100k background traces. Even with the Tor HS dataset, DNN classifiers performed effectively with 94-98% TPR and 3-4% FPR in multiclass setting.

In particular, in the binary classification, both classifiers presented almost 0% FPR, indicating that WF attacks based on DNN models become more feasible since they reduce the chances of misclassification.

### 6.2 Keyword Traffic Analysis

Table 5: KF with MLP(M), and CNN(C) using both RESP and cell traces. ((b):binary classification, (m):multiclass classification)

| Metrics | RESP(C) | RESP(M) | Cell(M) |
|---|---|---|---|
| TPR(b) | 86±1 | 88±1 | 59±2 |
| FPR(b) | 5 | 5±1 | 11±2 |
| BDR(b) | 95 | 95±1 | 85±2 |
| WMacc(m) | 22±1 | 27±1 | 22±1 |
| BDR(m) | 59±2 | 63±1 | 50±1 |

We used 100 instances of each of 100 monitored keywords and 10,000 background keyword traffic instances in the KTT dataset and two feature sets, RESP traces and Tor cell traces.

*Features.* Since Oh *et al.* [26] proposed that RESP-based features make KF attacks stronger, we explored whether this is true with MLP classifiers. Compared to Tor cell traces, Table 5 shows that RESP features substantially enhanced the performance of MLP classifiers. In addition, with RESP features, we achieved better performance (88% TPR and 5% FPR) than svmResp in the binary classification since svmResp yielded 82.6% TPR and 8.1% FPR.

*Top K analysis.* We evaluated DNN classifiers using WMacc, as used in Oh *et al.* [26]'s work. Compared to binary classification, multinomial classification results using DNN classifiers were not as powerful, as shown in (Table 5). To further investigate the performance of multiclass classifiers, we adopted the top *k* analysis by considering additional predicted labels that yield reasonably high prediction probabilities.

As shown in Figure 4, top *k* analysis helps to substantially improve the quality of multiclass classification for both Tor cell and RESP features. When using the top 5 analysis, MLP classifiers derived higher WMacc than svmRESP (62% vs. 55%).

*Confidence threshold and network.* As shown in Section 6.1, increasing the confidence threshold deteriorates the performance of DNN classifiers. Although overall CNN classifiers are more robust against this factor, as shown in Figure 5b, they also exhibit larger standard deviation with high confidence thresholds than MLP classifiers. There is much less distinction power between keyword traces, and in particular fewer local patterns to be learned by filters, making CNNs a less ideal choice for this task. However, with more training data the prediction results become more stable across different portions of the dataset.

*Summary.* KF using DNN classifiers outperformed svm-Resp [26], and using the top-*k* approach improves the multinomial classification capability of MLP classifiers for KF attacks.

### 6.3 TLS-encrypted Traffic Analysis

| Metrics | packet time | TCP packet | TLS record |
|---------|-------------|------------|------------|
| TPR (%) | 83.03 | 81.74 | 82 |
| FPR (%) | 9.55 | 10.64 | 12.53 |

Table 6: MLP performance over TLS-encrypted traffic data.

In this section we evaluate DNN classifiers using TLS-encrypted traces by varying feature set and dimensions. We used the WST dataset, consisting of 9000 monitored instances from Alexa top 100 websites and 9000 background website instances.

*Features.* We built three types of sequences, based on inter-packet timing, the size of TCP packets, and the length of TLS records. In the first trial, TCP packet and TLS record sequences performed poorly (For TCP, 48% TPR and 11% FPR, for TLS, 50% TPR and 14% FPR). Using MLP models, we achieved the best performance, 83% TPR and 10% FPR, using the sequence of packet directions, giving the results shown in Table 6.

*Network.* We further evaluated CNN classifiers using the representation based on the sequence of TLS record sizes, and achieved 92% TPR and 4% FPR. CNN classifiers improved on the result obtained by MLP classifiers (92% vs. 82%). Filters in CNN produce stronger representations of the step by step interactions in website downloads based on the local input patterns in TLS-encrypted traffic.

*Summary.* MLP classifiers performed better with the feature sequence consisting of binary information, -1 and 1. This transformation is consistent with the intuition behind batch normalization [17], which leads to higher learning rates without sacrificing the accuracy by solving internal covariate shifts. With the input vectors consisting of -1 and 1, MLP classifiers are less prone to high gradients, which could result from large TLS records.

### 6.4 Defended Traffic Analysis

In this section, we evaluated DNN classifiers against recent WF defenses, BuFLO [11], Tamaraw [10], WTF-PAD [18], and Walkie-Talkie [40]. BuFLO [11] pads dummy packets to fill in timing gaps and further extends the transmission to send packets of fixed length at fixed intervals. Tamaraw [10] improves BuFLO to be a more efficient and effective defense by using different padding intervals for incoming and outgoing packet direction and fixing outgoing packets at a higher packet interval, which reduces the overhead for infrequent outgoing traffic. To propose lighter-weight WF defenses, Juarez et al. [18] focused on hiding large gaps between bursts, which make traffic instances distinct, and randomly selected gaps and filled them in to hide this traffic pattern. WTF-PAD significantly decreases bandwidth overhead as well as latency while yielding good performance against *k*-NN classifiers. Walkie-Talkie [40] is an efficient WF defense technique based on half-duplex communication and burst molding, which make many packet sequences the same and add fake cells to burst sequences to be molded into the supersequence.

*BuFLO and Tamaraw.* Due to the bandwidth overhead, feature vectors for these data sets had higher dimension: for MLP classifiers, we used 20,164 (BuFLO) and 15,129 (Tamaraw) features and for CNN classifiers, we generated 30,000 (BuFLO) and 25,000 (Tamaraw) features. For Table 7, we evaluated DNN classifiers using Wang and Tor HS dataset in the closed-world setting, following Hayes and Danezis [13]. Based on Table 7 and Table 1 reported by Hayes and Danezis [13], MLP classifiers performed effectively the same as *k*-FP classifiers did against BuFLO and much better than other WF attacks under Tamaraw. CNN classifiers outperformed all other classifiers against both defenses, yielding 46-53% accuracy against BuFLO and 17-35% against Tamaraw.

*Walkie-Talkie.* We used the dataset provided by Wang and Goldberg [40], to train and test DNN classifiers. As shown in Table 1 [40] and Table 8, even though Walkie-Talkie reduced the accuracy of MLP classifiers, MLP classifiers still worked adequately and the top-$k$ approach somewhat improved the performance of MLP classifiers. More surprisingly, CNN classifiers yielded much better quality of multinomial classification with accuracy of 54-65%.

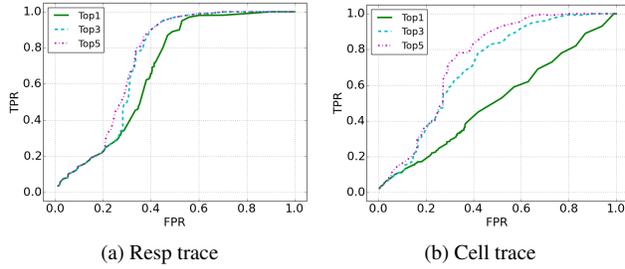
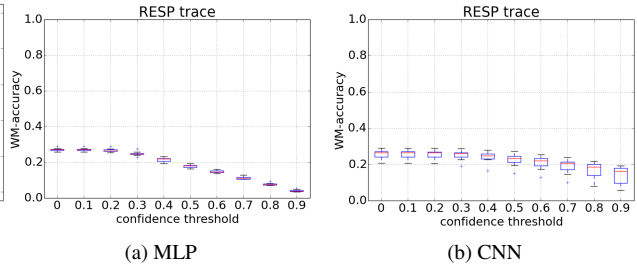

| (a) Resp trace | (b) Cell trace | (a) MLP | (b) CNN |

Figure 4: ROC of Top-K analysis on KF for multiclass classification

Figure 5: KF(RESP) with MLP and CNN by varying the confidence threshold

Table 7: MLP and CNN performance under BuFLO(B) and Tamaraw(T) (Non:No defense).

| DNN | MLP | | | | | | CNN | | |
|---|---|---|---|---|---|---|---|---|---|
| Dataset | Wang | | | TorHS | | | Wang | | |
| Defense | Non | B | T | Non | B | T | Non | B | T |
| Bandwidth overhead(%) | 0 | 217 | 181 | 0 | 1,240 | 1,019 | 0 | 217 | 181 |
| Top1(%) | 86±2 | **19±1** | **15±2** | 60±2 | **34±1** | **14±1** | 92±1 | **45±4** | **17±1** |
| Top3(%) | 91±2 | **35±2** | **26±2** | 67±2 | **48±2** | **23±1** | 96 | **53±2** | **33±2** |
| Top5(%) | 91±2 | **38±3** | **27±3** | 68±2 | **50±2** | **23±1** | 96 | **53±2** | **35±2** |

Table 8: The performance of DNN classifiers under Walkie-Talkie. For Top $k$ analysis, we chose the confidence threshold leading optimal accuracy. Note that Undef means traces, collected without the defense and Def indicates traces, collected under the defense (Bandwidth overhead=113%).

| | MLP | | CNN | |
|---|---|---|---|---|
| Top$k$ | Undef | Def | Undef | Def |
| Top1 | 84±5 | **26±3** | 89±3 | **54±1** |
| Top3 | 88±3 | **37±4** | 93±3 | **64±1** |
| Top5 | 89±3 | **38±3** | 93±3 | **65±1** |

Table 9: The performance of CNN classifiers under WTF-PAD. For Top $k$ analysis, we chose the confidence threshold yielding optimal accuracy (For Juarez, bandwidth overhead=166%, for WTF, bandwidth overhead=174%).

| | Juarez(34) | | WTF(100) | WTF(550) |
|---|---|---|---|---|
| Top$k$ | Undef | Def | Def | Def |
| Top1 | 88±1 | **53±1** | 53±1 | 58±1 |
| Top3 | 93±1 | **57±1** | 58±1 | 65±1 |
| Top5 | 94±1 | **58±1** | 58±1 | 65±1 |

*WTF-PAD.* We explored CNN classifiers using two datasets, one provided by Juarez *et al.* [18] (named Juarez dataset) and the other shared by Li *et al.* [23], which we call the WTF dataset. Juarez dataset comprises 34 instances of each of 100 websites and WTF dataset includes 100-600 instances of the Alexa top 100 websites. As shown in Table 9, even with 34 instances, CNN classifiers presented reasonable closed-world accuracy, 53-58%. Based on WTF dataset, they performed much better with more training dataset using 550 instances.

*Summary.* MLP classifiers performed adequately against WF defenses, however, CNN classifiers showed superior accuracy against all 4 defenses. The superior capability of CNNs against Walkie-Talkie and WTF-Pad show that defenses focused on burst padding are not enough to hide local patterns against filters in CNNs even though they ensure more efficient bandwidth overhead.

### 6.5 Recent WF Attacks with AE Features

We revisited cutting edge WF attacks [13, 28, 38] and used the machine learning algorithms chosen in those works, with new features, learned by an AE (AE features). Feature engineering, suggested by recent WF research [13, 28], usually requires a considerable amount of human efforts such as manually inspecting the traffic pattern and experimentally deciding the optimal dimension of feature vectors. Feature extraction based on AEs offers the potential to automate this process. We used 90 instances of each of 100 monitored websites and 9,000 background websites in Wang's dataset.

Table 10: Performance of state-of-the-art machine learning algorithms with AE features (dim: feature dimension).

| | Binary | | | | Multiclass | | | |
|---|---|---|---|---|---|---|---|---|
| dim | 10 | | 80 | | 20 | | 80 | |
| | TPR | FPR | TPR | FPR | TPR | FPR | TPR | FPR |
| $k$-NN | 96±1 | 3±1 | 98±1 | 2±1 | 82±2 | 2±1 | 84±2 | 2±1 |
| SVM | 95 | 3 | 98 | 1 | 97 | 12 | 97 | 10 |
| $k$-FP | 94±1 | 2±1 | 96±1 | 1±1 | 81±2 | 1±1 | 83±1 | 1±1 |

Table 11: The best performance using Wang's dataset (AE(n): n features).

| Metrics | TPR | FPR |
|---|---|---|
| CUMUL | 96.59 | 9.55 |
| $k$-NN | 90.17 | 10.26 |
| $k$-FP | 88 | 0.5 |
| SDAE | 92.42 | 12.54 |
| AE(80)+$k$-NN | 97.89 | 2.07 |
| AE(100)+SVM | 97.6 | 1.47 |
| AE(80)+$k$-FP | 96.2 | 0.98 |
| MLP | 96.56 | 2.78 |
| CNN | 96.72 | 1.78 |

To feed AE features into WF classifiers, we adopted the implementation of CUMUL [28] and $k$-FP [13] after removing feature extraction part. However, since the implementation of $k$-NN [38] is suited to features based on a Tor cell trace, we implemented $k$-NN using `sklearn.neighbors` with weight learning that computes the inverse of the distance between neighbors to handle AE features in a better manner.

To understand how the dimension of AE features impacts the performance of classifiers, we varied the number of units, 10-100, in the hidden layer, which encoded website traffic vectors. This layer corresponds to the blue layer in Figure 8c in Appendix B. In implementation perspective, this procedure is very simple since we can save a trained AE model at the end of training and extract compressed features on target data using weights of an encoder network, kept in a trained model.

Since the dimension of this hidden representation is significantly lower than the length of the original input vector, this leads to dimensionality reduction. After we performed the open-world evaluation based on Wang's dataset, we realized that the dimension of hidden units of an AE scarcely impacted the performance of machine learning algorithms in binary classification. Similarly, in multiclass classification, there was almost no change on both metrics for feature vectors with dimension larger than 20 (Figure 10 in Appendix E). As shown in Table 10, AE features enabled classifiers to yield comparable performance to state-of-the-art WF attacks [13, 28, 38], even with much lower feature dimensionality and no human effort. We achieved much lower FPR than $k$-NN [38] and higher TPR than $k$-FP [13] while spending much shorter time in feature engineering. By the comparison to CUMUL [28], we obtained similar results in much shorter training time since classifier learning with AE features of dimension 20 took 39 minutes while the same process with CUMUL features took 4 hours. Note that training time does not include the time spent on feature extraction.

In Table 11, we reproduced state-of-the-art WF attacks [13, 28, 38] and Stacked Denoising Auto Encoder (SDAE) [31] using Wang's dataset in open-world setting. For SDAE, we discovered the optimal parameters for our experiment setting by running hyperparameter tuning over the search space, reported in the paper [31] and used them for training SDAE. AE features improved the performance of all three attacks and MLP and CNN classifiers presented comparable performance without the separate feature extraction procedure.

*Summary.* Based on these experiments, DNN's ability to automate feature extraction significantly reduces effort required for feature engineering with other classification algorithms as well as generating cost-effective features. In addition, the dimensionality reduction is another advantage to ensure faster running time of algorithms because AE keeps as much information as it can in a compressed representation and these lower-dimensional features reduce the training time of classifiers. Furthermore, since the cost function plays an important role in training, it impacts the encoded data, which indicates that by adjusting the cost function with in-depth understanding on the input data, we can elaborate this feature engineering procedure to generate more powerful features.

### 6.6 Fingerprintability Prediction

Webpages exhibit a range of "fingerprintability," in that some pages are easy to identify by their traces while others are not. To study what factors contribute to this property, we used the following methodology. First, we compute the *accuracy* of each website by training and testing an MLP classifier using 90 instances each of the Alexa top 50 web sites, and single instance of 9000 back-

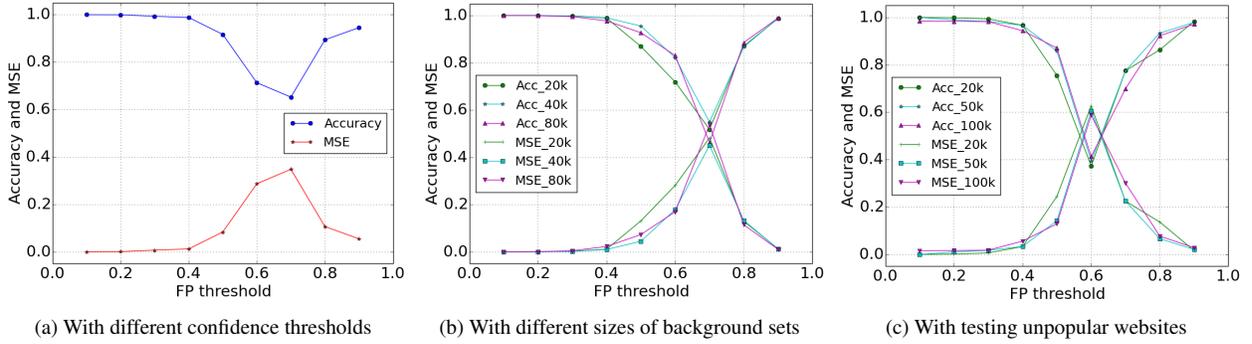

Figure 6: Accuracy and MSE when varying the FP threshold and the size of background dataset

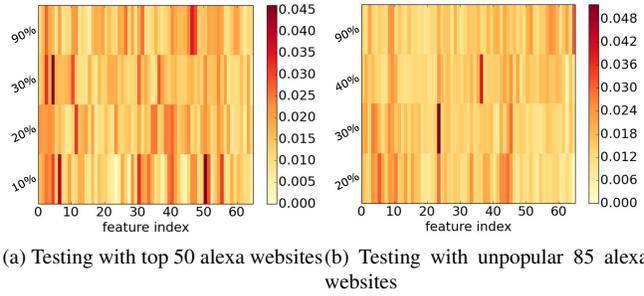

(a) Testing with top 50 alexa websites (b) Testing with unpopular 85 alexa websites

Figure 7: The feature importance of each of 65 features according to FP thresholds.

ground websites. Note that we used 10 iterations, where each iteration randomly selects training and testing data with the ratio 60:40. Then, based on prediction results, we computed the fraction of instances labelled as True positives for the accuracy. These websites were used as "training" data for the fingerprintability (FP) classifier. We followed the same procedures with Alexa top 61-100 monitored websites to determine the *accuracy* of the "testing" websites.

Based on HTML document files for each training and testing websites, we constructed 65 *HTML features*, shown in Appendix D, computed ranks for each feature, and labeled HTML feature vectors with 1 or 0, depending on whether the corresponding website *accuracy* was greater than a certain FP threshold (In other words, the website is *fingerprintable*) or not, respectively. Finally, we trained MLP classifiers as FP classifiers with those features and labels.

***FP threshold.*** To define the fingerprintability of each website traffic instance, we varied *accuracy* thresholds, 10-90%, to show how increasing the threshold impacts the performance of classifiers.

As shown in Figure 6a, if we increased the FP threshold, the accuracy of classifiers continually decreased and the mean squared error (MSE) increased until the threshold reached 70%. After that, the accuracy was increasing with decreasing MSE. According to Figure 6, we achieved accuracy of 98-99% for 10-40% thresholds and 95% for 90% threshold. This shows that the FP classifiers performed better with relatively smaller or higher thresholds. Section 5.2 described how we applied class weights to derive both the accuracy and the error rate.

***Feature importance.*** We adopted Gini importance, explained in Section 5.2, to derive the feature importance score and Figure 7 shows those scores for different confidence thresholds. Then, we computed the sum of those scores for each of 65 features to select the top 15 features, shown in Table 13 of Appendix F.

The total number of unique domains (feature index 5) and third party links (feature index 3) are more important HTML features for Alexa top 61-100 websites and this is consistent with our expectation that the traffic regarding the communication with third party websites ruins traffic pattern to be indistinguishable (Table 12). The number of characters and words and embedded web objects such as images (feature index 30,40,50-55), included in HTML document files, are also important features. Table 12 indicates that web pages carrying less amount of static web contents (e.g., texts, images,etc.) are easy to fingerprint while webpages including dynamic web objects (e.g., html) [3] are hard to fingerprint.

All these top features highly affect the fingerprintability across different FP thresholds as shown in Figure 7a.

***Background set.*** To show the effect of the amount of background data on the performance of FP classifiers, we used 20,000, 40,000, and 80,000 background website instances. According to Figure 6b, feeding more unmonitored traffic instances had little effect on the accuracy and furthermore, the performance of FP classifiers was almost the same against 10-40% FP thresholds. Overall, the worst accuracy in FP predictions resulted from using a FP threshold of 70%, thus we suggest closer numbers (e.g.,10,20,90%,etc.) to either end in the accuracy range for the FP threshold to produce the best results.

---
[3] We found that most dynamic web contents such as news and feeds are embedded in a webpage with a file extension, `html`.

Table 12: Important features for fingerprintable (accuracy $\geq$ 90%), and less fingerprintable (accuracy $\leq$ 10%) websites.

| Feature Index | 3 | 5 | 30 | 40 | 51 | 55 |
|---|---|---|---|---|---|---|
| 90% | 16±10 | 9±6 | 28k±25k | 16k±15k | 196±125 | 23±19 |
| 10% | 109±37 | 49±16 | 14k±5k | 3k±218 | 12±10 | 459±157 |

*Unpopular websites.* To investigate the impact of different testing dataset on FP classifiers, we furthermore collected 90 traffic instances of each of 85 unpopular Alexa websites, ranked around 1 million, and downloaded their HTML document files. We trained MLP using HTML features of Alexa top 67 websites (90 instances for each) and tested it with HTML features of 5,100 unpopular website traffic instances. In this experiment, we show how FP classifiers perform toward unpopular websites.

For good FP thresholds (10-40% and 90%) identified in previous experiments, we achieved slightly higher error rates while the accuracy was almost the same (Figure 6c). For worse FP thresholds (60-70%), MSE became much greater and the accuracy also decreased. That is, across different sizes of background datasets, the fingerprintability prediction on unpopular websites has more uncertainty with increased error rates and this becomes more severe with bad FP thresholds. However, 10-30% and 90% FP thresholds still yielded trustworthy results with 98-99% accuracy and MSE of 0.1-2%.

We listed the top 15 features based on feature importance scores for HTML features of 85 unpopular websites in Table 14 of Appendix F. As shown in Figure 7b, even though different features appeared in the top list, statistics about domains (feature index 4-5) and embedded web contents (feature index 43) still impact the fingerprintability of unpopular websites, which is also consistent with previous results using Alexa top websites.

Moreover, we investigated feature 4-5 and feature 43 to observe distinction between popular and unpopular websites. The degree of variation in all three features becomes higher in popular websites and more unique domains (7 vs. 4) and embedded web objects are included in popular websites (67 vs. 45). This leads increased MSE while weakening the performance of FP classifiers when we evaluated against unpopular websites.

*Labeling with results of k-FP.* In this experiment, we show that our FP classifiers can be applicable to predict fingerprintability against other state-of-the-art WF attacks. To generate training labels, we ran *k*-FP [13] classifiers with our dataset consisting of Alexa top 67 monitored websites and 20,000 background dataset. To obtain testing labels, we applied the same experiment setting with 82 unpopular Alexa websites. Then, we extracted 65 HTML features based on elements of their HTML documents and based on results, returned by *k*-FP classifiers, we computed the accuracy for each monitored website and labelled features with 1 or 0 in a way we described earlier. Finally, we trained MLP classifiers with 67 popular website and tested it with 82 unpopular websites. With 30% FP threshold, we achieved 98% of accuracy and 2% MSE. This indicates that our fingerprintability classifiers can be further deployed for WF defenses against recent WF attacks. We will leave further investigation on other WF attacks for future works.

*Summary.* Based on FP classifiers, we can guide developers with safer design of websites against traffic analysis. Furthermore, we can develop a fuzzing-like tool to 1) automatically scrutinize documents to identify vulnerable patterns in HTML DOM, which makes website traffic more distinct, 2) propose more resilient HTML DOM, and 3) test the new DOM to check if it actually minimizes the predicted fingerprintability. This can prevent the website from leaking the web visiting activity of the user or endangering the onion service location.

## 7 Conclusion

We extensively explored DNNs for traffic analysis and proposed 3 different applications, automated feature engineering, fingerprinting attacks, and fingerprintability prediction. As a feature extractor, lower dimensional representations, learned by an AE, made state-of-the-art WF attacks more effective as well as efficient. For fingerprinting attacks, DNN performed nicely across various traffic datasets and different fingerprinting tasks, and against recent WF defenses. Lastly, MLP classifiers successfully decided the fingerprintability of websites using HTML features, leaving the future work on WF defenses using FP classifiers to ensure safer onion service.

# A  Appendix

# B  Deep Neural Network

In this section, we briefly discuss Multilayer Perceptron (MLP), Convolutional Neural Network (CNN), which are methods for supervised learning, and Autoencoder (AE), which is an unsupervised learning method.

*Multilayer Perceptron.* MLP [12, 32], shown in Figure 8a, is a basic neural network, a sort of feed forward network, and also is known as a backpropagation algorithm. It consists of an input layer, one or more fully-connected hidden layers and an output layer. Therefore, MLP always has at least 3 layers. In fully-connected layers, all nodes have full connections to all activations in previous layer. Activation functions are computed by a matrix multiplication, which is followed by a bias offset.

MLP two procedures, *forward propagation*, which initializes weights and forwards pass through multiple layers to produce the output, *back propagation*, which calculates errors of the output layer, and then updates weights layer by layer. A single pass through is called an *epoch* and consists of multiple *batches*.

We applied the softmax function in the output layer, which is the generalization of the binary Logistic Regression to multiclass setting. It takes the vector of arbitrary real values and a vector of values in [0,1], where the sum is 1. This real-valued score is normalized class probability. Since we used a cross entropy to compute the loss, we analyze it as unnormalized log probability for each class and apply a cross entropy loss, $E = -\sum_{i}^{nClass} t_i log(y_i)$, where $i$ is a class index, *nClass* is the total number of classes, $t_i$ is a target probability, and $y_i$ is an output probability. The total loss for the dataset is computed by the mean of $E$ over all training samples.

*Convolutional Neural Network.* CNN [21], shown in Figure 8b, consists of one or more convolutional layers, followed by one or more fully connected layers. The forward propagation runs three series of operations, *convolution*, *pooling*, and *classification*. The convolution operation extracts features from the input by learning features using small squares of input, which are called *filters* or *kernels*. That is, we slide filters across the width and height of the input and calculate dot products between entries of the filter and the input to generate a 2D feature map. Sliding different filters over the same feature generate different feature maps and CNN can learn meaningful pattern through this procedure.

Pooling reduces the dimension of the feature maps and thus, the amount of parameters and computation in the network. Max pooling layer operates on selecting the max element from the feature map for resizing spatially. Then, high level features learned by convolutional and pooling layers are fed into MLP for the classification. For the back propagation, it keeps track of the index of max activation so that routing the gradient becomes simpler than general backward pass.

*Autoencoder.* AE [5], shown in Figure 8c, is unsupervised neural network, which consists of two neural networks, an *encoder*, which learns lower-dimensional data abstractions, and a *decoder*, which recovers the original data. It aims to predict the input by using less number of hidden neurons than input nodes to learn as much information as it can learn to hidden neurons. More specifically, since the number of hidden nodes at each hidden layer is less than the dimension of the original input vector, the network is forced to learn a compressed representation of the input data and then reconstruct the input. Through these procedures, the network can discover interesting structure of the data.

One advantage with an AE is that at the end of training, we can have weights that lead to the hidden layer (blue layer in Figure 8c), we can train using certain input. Furthermore, when we meet other data later, we can reduce its dimensionality using those weights without retraining. âĂŁ Thus, it provides benefits to reduce the feature dimensionality for data visualization and reducing the noise in the data. Hidden units in an encoder keep as much information as it can while denoising the data. The attractive part is that we can elaborate feature extraction using encoded data through the cost function since we have a lot of choice on the cost function and can adjust the weight for each class and sample. We can use this power to reflect certain phenomenons in the dataset, which eventually leads to a more efficient and meaningful data representation. In Section 6.5, we focus on this functionality of dimensionality reduction [15]. We used MLP for an *encoder* and a *decoder*, while varying the number of *hidden units* in a hidden layer of an encoder.

To construct a more generative model, Variational AE (VAE) was introduced by Kingma and Welling [19] and Rezende *et al.* [30]. Instead of memorizing a fuzzy data structure, it generates latent vectors following a Gaussian distribution by forcing a constraint to an encoder. Subsequently, to compute the loss of a VAE, two types of losses must be considered, the error between the input and reconstructed data, and the loss between latent variables and a unit Gaussian, reflected by KL divergence. Training VAE is tricky due to the trade off between these two different losses. Improvement in the generalization also promotes the quality of data reconstruction by a decoder.

*Avoid overfitting.* DNN usually struggles with overfitting, which means that the network memorized the training samples and hence, the error in testing dataset is large even though the training loss is tiny. To overcome this issue, dropout [36] and regularization are widely used. The dropout temporarily removes units in layers based on the probability of each unit to be retained. Regularization is also known as weight decay, which means that it penalizes large weights based on constraints on their squared values (L2) or absolute values (L1). We applied these techniques to both MLP and CNN architectures in Section 4.4.

## C  LRP Score and Tor Cell Direction.

For each Tor cell direction, -1(incoming), 0(padding), and 1(outgoing), we scatter LRP scores in Figure 9. As shown in the graph, there is no distinction among incoming, outgoing, and padding Tor cells and the average of each of them is 0.001 and the standard deviation is very small enough to be negligible.

## D  HTML Features

We list 65 features, extracted based on an HTML DOM and offer details how we generated those in Section 5.1.

- 1. total number of links
- 2. total number of links from same domain
- 3. total number of third party links
- 4. total number of domains in links
- 5. total number of unique domains in links
- 6. total number of tag paths
- 7. total number of unique tag paths
- 8. sum of number of unique tags per a path
- 9. median of number of unique tags per a path
- 10. mean of number of unique tags per a path
- 11. std of number of unique tags per a path
- 12. total number of change of tag path direction (if depth increases, positive, otherwise, negative)
- 13. total number of non change of tag path direction
- 14. total number of positive direction in tag paths
- 15. total number of negative direction in tag paths
- 16. total sum of tag depths
- 17. std of tag depths
- 18. total number of max depth in tag paths
- 19. total number of min depth in tag paths
- 20. total number of median depth in tag paths
- 21. total number of rounded mean depth in tag paths
- 22. total number of 30% percentile of depth in tag paths
- 23. total number of 70% percentile of depth in tag paths
- 24. max depth of tag paths
- 25. min depth of tag paths
- 26. median depth in tag paths
- 27. rounded mean depth in tag paths
- 28. 30% percentile of depth in tag paths
- 29. 70% percentile of depth in tag paths
- 30. total number of tags
- 31. total number of unique tags
- 32. total number of comments
- 33. total number of attributes
- 34. total number of unique attributes
- 35. total number of characters
- 36. total number of characters in script tag
- 37. total number of characters in style attribute
- 38. total number of characters in attribute
- 39. total number of characters in data including those in script and style attributes
- 40. total number of characters in data attribute
- 41. total number of words in data including those in script and style attributes
- 42. total number of words in data attribute
- 43. total number of image tags in HTML DOM
- 44. portion of image tags over all tag paths
- 45. total number of png files in HTML DOM

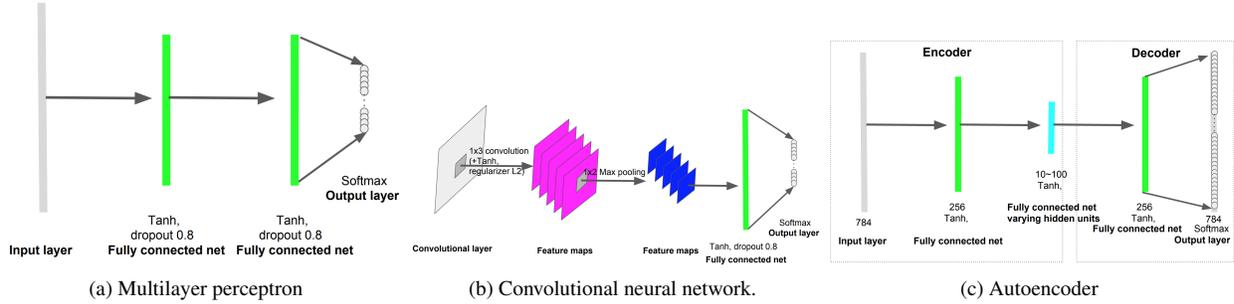

(a) Multilayer perceptron  (b) Convolutional neural network.  (c) Autoencoder

Figure 8: Our architectures on 3 DNN models.

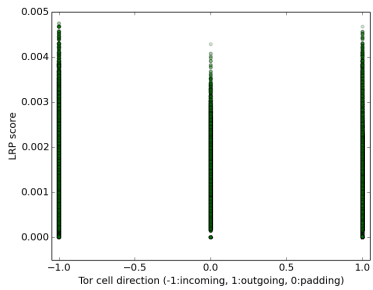

Figure 9: LRP scores depending on Tor cell direction.

- 46. portion of png files in HTML DOM
- 47. total number of ico files in HTML DOM
- 48. portion of ico files in HTML DOM
- 49. total number of jpg files in HTML DOM
- 50. portion of jpg files in HTML DOM
- 51. total number of gif files in HTML DOM
- 52. portion of gif files in HTML DOM
- 53. total number of bmp files in HTML DOM
- 54. portion of bmp files in HTML DOM
- 55. total number of html files in HTML DOM
- 56. portion of html files in HTML DOM
- 57. total number of css files in HTML DOM
- 58. portion of css files in HTML DOM
- 59. total number of js files in HTML DOM
- 60. portion of js files in HTML DOM
- 61. total number of mp3 files in HTML DOM
- 62. total number of avi files in HTML DOM
- 63. total loading time in a capture (pcap) file
- 64. size of a html document file
- 65. size of a capture file

## E  The Relationship between AE Feature Dimension and Performance of Traditional ML Techniques.

We varied the number of nodes in the second hidden layer (Figure 8c) to show its impact on the performance of state-of-the-art machine learning algorithms, SVM, $k$-NN, and $k$-FP classifiers. As shown in Figure 10, AE feature dimension gave almost no change in the quality of binary classification and the multiclass classifiers performed similarly with AE features whose dimension is greater than 20. This indicates that AE has superior ability to capture meaningful pattern about the input data while reducing the feature dimensionality.

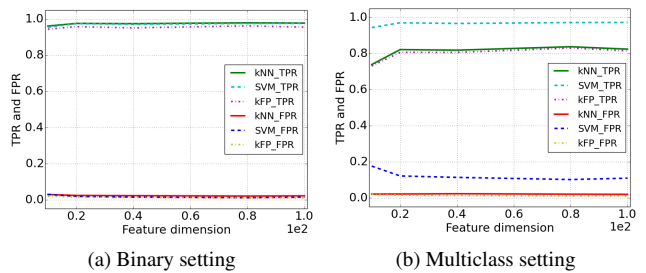

(a) Binary setting  (b) Multiclass setting

Figure 10: Feature importance score using LRP.

## F  Top 15 Features for Fingerprintability Prediction over Unpopular Websites

We computed the sum of feature importance score for each of 65 HTML feature indices and selected top 15

Table 13: Top 15 html properties based feature importance measured by gini index

| Top 15 Features |
|---|
| total number of unique domains in links |
| total number of third party links |
| total number of unique tags |
| total number of gif files attached in document |
| total number of html files attached in document |
| total number of words in data including script and style |
| proportion of html files attached in document |
| total number of ico files attached in document |
| total number of unique tag paths |
| portion of ico files attached in document |
| total number of domains in links |
| std of number of unique tags per a path |
| proportion of gif files attached in document |
| total number of change of tag path direction |
| total number of characters in data field |

features, which are more important to predict the fingerprintability. Top 15 features to predict the fingerprintability for Alexa top 61-100 websites are listed in Table 13 and those for Alexa 85 unpopular websites are listed in Table 14. As mentioned in Section 6.6, features based on domains and embedded web contents impact the fingerprintability of websites across different datasets.

Table 14: Top 15 html properties based feature importance measured by gini index when using 85 unpopular websites

| Top 15 Features |
|---|
| maximum depth of tag paths |
| median of number of unique tags per a tag path |
| mean of number of unique tags per a tag path |
| total number of domains in HTML DOM |
| total number of png files attached in HTML DOM |
| total number of tag paths in HTML DOM |
| proportion of image tags over all tag paths |
| total number of image tags in HTML DOM |
| total number of unique domains in links |
| total number of unique attributes in HTML DOM |
| std of number of unique tags per a tag path |
| std of number of unique tags per a tag path |
| total number of characters in data including script and style attributes |
| total number of links in HTML DOM |
| proportion of gif files attached in HTML DOM |